\documentclass[%
reprint,
superscriptaddress,
showpacs,preprintnumbers,
 amsmath,amssymb,
 aps,
prb,
]{revtex4-1}

\usepackage{graphicx}
\usepackage{dcolumn}
\usepackage{bm}
\usepackage{color}

\begin{document}


\title{Infrared magneto-spectroscopy of graphite in tilted fields}
\author{N.\ A.\ Goncharuk}
\altaffiliation{present address: ABB s.r.o., Semiconductors, R\&D,
  Novodvorsk\'{a} 1768/138a, 14221 Prague, Czech Republic}
\affiliation{Institute of Physics,
Academy of Science of the Czech Republic,~v.v.i.,\\
   Cukrovarnick\'{a} 10, 162 53 Prague 6, Czech Republic}
\author{L. N\'advorn\'{\i}k}
\affiliation{Institute of Physics,
Academy of Science of the Czech Republic,~v.v.i.,\\
   Cukrovarnick\'{a} 10, 162 53 Prague 6, Czech Republic}
\affiliation{Charles University in Prague, Faculty of Mathematics and Physics, 
  Ke Karlovu 3, 121 16 Praha 2, Czech Republic}
\author{C.\ Faugeras}
\affiliation{Laboratoire National des Champs Magn\'{e}tiques Intenses,
  CNRS-UJF-UPS-INSA, 25, avenue des Martyrs, 38042 Grenoble, France}
\author{M.\ Orlita}\email{orlita@karlov.mff.cuni.cz}
\affiliation{Charles University in Prague, Faculty of Mathematics and Physics, 
  Ke Karlovu 3, 121 16 Praha 2, Czech Republic}
\affiliation{Laboratoire National des Champs Magn\'{e}tiques Intenses,
  CNRS-UJF-UPS-INSA, 25, avenue des Martyrs, 38042 Grenoble, France}
\author{L.\ Smr\v{c}ka}\email{smrcka@fzu.cz}
\affiliation{Institute of Physics,
Academy of Science of the Czech Republic,~v.v.i.,\\
   Cukrovarnick\'{a} 10, 162 53 Prague 6, Czech Republic}

\date{\today}

\begin{abstract}
The electronic structure of Bernal-stacked graphite subject to tilted
magnetic fields has been investigated using infrared magneto-transmission experiments.
With the increasing in-plane component of the magnetic field $B_\|$,
we observe significant broadening and partially also splitting of interband inter-Landau
level transitions, which originate at the $H$ point of the graphite Brillouin zone,
where the charge carriers behave as massless Dirac fermions.
The observed behavior is attributed to the lifting of the twofold degeneracy
of Landau levels at the $H$ point -- a degeneracy which in graphite complements the 
standard spin and valley degeneracies typical of graphene.
\end{abstract}

\pacs{71.20.-b, 71.70.Di}
\maketitle
\section{Introduction}
It was the fabrication of single-layer
graphene\cite{BergerJPCB04,NovoselovScience04} and subsequent
discovery of massless Dirac fermions\cite{geim05,Kim_2005} which
triggered the present increased interest in the electrical and optical
properties of graphite\cite{LiPRB06,LukyanchukPRL06,GonzalesPRL07,OrlitaPRL08,OrlitaJPCM08,OrlitaPRL09,ChuangPRB09,ZhuNP10,Ubrig_2011,Tung_2011,KossackiPRB11,KimPRB12,LevalloisSSC12}
-- supposedly a well-known material for the condensed matter physics.

Even though graphene is a purely two-dimensional (2D) system and
graphite is characterized by a (highly anisotropic but still) clearly
3D band structure, these materials, as demonstrated experimentally,\cite{OrlitaPRL08,OrlitaPRL09,OrlitaSSC09}
share surprisingly similar optical response when the magnetic field
$\mathbf{B}$ is applied perpendicularly to layers.  A simple model, invoking
inter-Landau level excitations between highly-degenerate Landau levels
(LLs) of massless Dirac fermions, implies the magneto-optical response
that is linear in $\sqrt{B}$, see, \textit{e.g.}, Refs.~\onlinecite{SadowskiPRL06,GusyninPRL07,JiangPRL07,DeaconPRB07,OrlitaPRL08II,CrasseeNP11,CrasseePRB11,BooshehriPRB12},
and is capable to account for a significant part of the
magneto-optical data acquired on graphite. Importantly, these data come
not only from recent magneto-transmission studies of thin
specimens,\cite{OrlitaPRL08,OrlitaPRL09,ChuangPRB09} but also
from original measurements carried out in late seventies,\cite{ToyPRB77}
in which $\sqrt{B}$-scaled spectral features have been observed using the
magneto-reflection technique. This pioneering work is a good candidate for
the first direct experimental observation of massless Dirac fermions, which
in bulk graphite coexist with massive particles and which provide more
conventional, \textit{i.e.}, linear in $B$ response.\cite{GaltPR56,SchroederPRL68}

The electronic band structure of graphite in the magnetic field is
mostly described using the standard model proposed by Slonczewski,
Weiss and McClure (SWM),\cite{S_W_1958, McClure_1960} even though
presumably more precise, but at the same time, also more
time-consuming approaches appeared recently, see, \textit{e.g.},
Refs.~\onlinecite{HoAPL11,HoPRB11}.  The SWM model has been derived in
late fifties using mostly symmetry arguments;  it describes the
electronic structure near the $H$-$K$-$H$ edge of the Brillouin zone
with energies not too distant from the Fermi level. The six of seven
parameters in the SWM model, $\gamma_0,\dots,
\gamma_5$,\cite{Brandt88} are usually interpreted as tight-binding
hopping integrals between the nearest-neighbor and partially also
next-nearest-neighbor atoms. An additional parameter $\Delta$, related
to the non-equivalence of carbon atoms in A and B positions, is
referred to as a pseudogap. All parameters must be considered rather
as adjustable parameters than true hopping integrals and are usually
obtained by fitting either experimental data or the results of {\em ab
  initio} calculations.\cite{Gruneis_2008} The importance of
individual parameters significantly varies, depending on the type of
experimental data for interpretation of which the SWM model is used.

For instance, the periods of Shubnikov-de
Haas\cite{WoollamPRL70,LukyanchukPRL04,LukyanchukPRL06,SchneiderPRB10}
(SdH) and de Haas-van Alphen\cite{WilliamsonPR65,SchneiderPRL12}
(dHvA) oscillations depend on the extremal cross sections of the
complex Fermi surface and all SWM tight-binding parameters must be
properly taken into account.  Similarly, cyclotron resonance
experiments,\cite{GaltPR56,DoezemaPRB79,OrlitaPRL12} which are also
sensitive to the immediate vicinity of the Fermi level and which provide fairly rich
response, can be hardly understood without the full SWM
model.

\begin{figure*}[t]
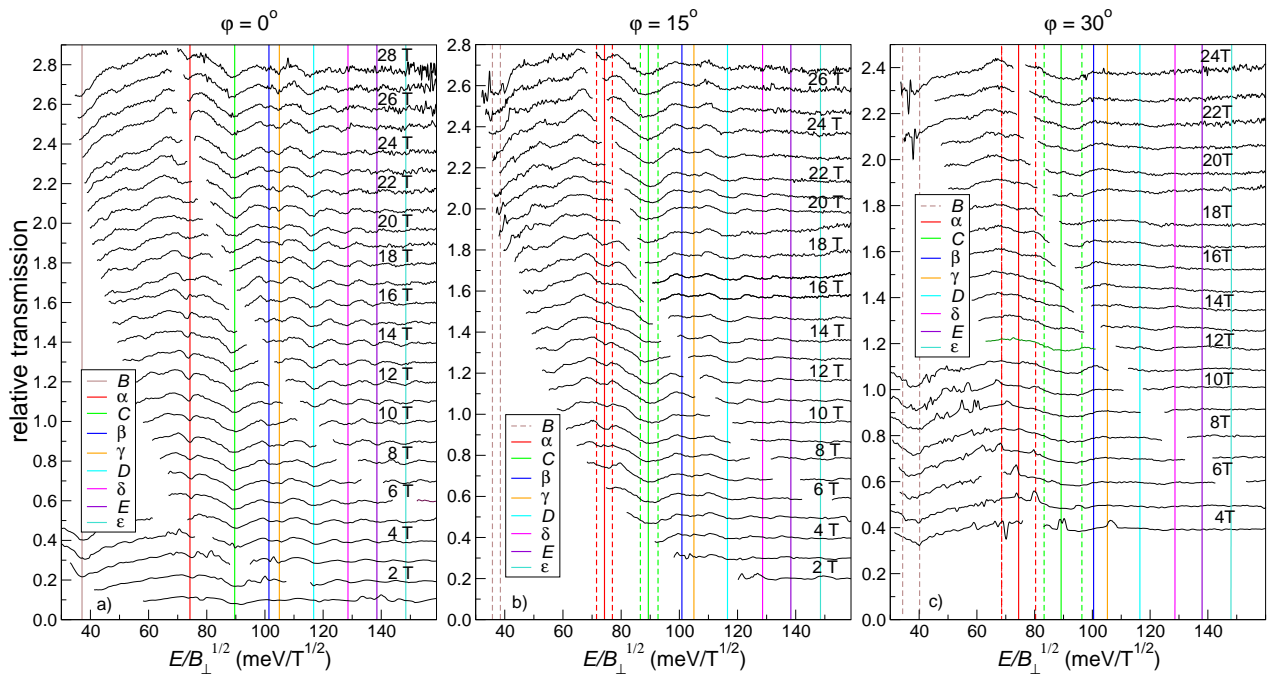

\begin{center}
\includegraphics*[scale=0.317]{fig1a.eps}
\includegraphics*[scale=0.317]{fig1b.eps}
\includegraphics*[scale=0.317]{fig1c.eps}
\end{center}
\caption{\label{Fig1} Transmission of a thin graphite specimen measured at magnetic field tilted by
$\varphi=0^{\circ}, 15^{\circ}$ and $30^{\circ}$ with respect to the $c$-axis of graphite. Values nearby individual
curves always denote component of the magnetic field perpendicular to the graphite layer $B_\perp$. The missing parts
of spectra correspond to regions in which the supporting tape is completely opaque.  The energy axis
for each curve is rescaled by a factor of $\sqrt{B_\perp}$ to facilitate identification of spectral features originating
at the $H$ point of graphite. Vertical lines correspond to positions of van Hove singularities in
the joint density of states as calculated using the minimal nearest-neighbor tight-binding model.\cite{Gonch_2012}
Since the lifting of degeneracy exactly at the $H$ point is governed by the coupling
constants $T_{n,n+1}\propto \sqrt{B_\perp} \tan(\varphi)$, the positions of these van Hove singularities
remain constant in these figures. For clarity, successive spectra in (a),(b), and (c) are shifted vertically by
0.1.}
\end{figure*}

On the other hand, interband transitions between electronic states far
away from the Fermi surface can be successfully described using a
simplified approach,\cite{OrlitaPRL09,ChuangPRB09} which models the
magneto-optical response of bulk graphite as a sum of responses of an
effective graphene bilayer and monolayer.\cite{Kos_2008} Notably, the
physical properties of a 3D system are thus described using responses
of two purely 2D materials, and interestingly, not more than two coupling
constants, intralayer $\gamma_0$ and interlayer $\gamma_1$, are needed
in the very first approach.\cite{PartoensPRB07} Within such a minimal model, the $H$ point
provides response similar to a single graphene sheet, but richer due to an additional twofold
degeneracy, and the $K$ point behaves as bilayer graphene, however,
with the interlayer coupling 
enhanced twice as compared to the true bilayer. Limits of this model
have been found, \textsl{e.g.}, by revealing the electron-hole
asymmetry at the $K$ point of bulk graphite in recent
magneto-transmission,\cite{ChuangPRB09}
magneto-reflection\cite{Tung_2011,LevalloisSSC12} and magneto-Raman
studies.\cite{KossackiPRB11,KimPRB12} The full SWM model has to be
used in such a case to get quantitative agreement between the
experimental data and theory.

In this paper, we set other limitations of the effective monolayer and bilayer
model for the magneto-optical response of graphite.
Namely, we test its applicability in experiments performed in the tilted-field configuration,
$\mathbf{B}=(0,B_{\|},B_\perp)$, which is a basic tool to distinguish between
2D and 3D character of condensed matter systems. We focus on the graphene-like signal
from the $H$ point and show that the magneto-optical response of graphite in a tilted
magnetic field follows the total magnetic field and not only its perpendicular
component, as should be in the case of an ideal 2D system. The infrared
magneto-transmission technique is thus, perhaps surprisingly,
significantly more sensitive to the in-plane component of the magnetic
field as compared to other techniques such as SdH or dHvA
oscillations, which reveal the 3D character of graphite only for rather
high tilting angles.\cite{SchneiderPRB10,SchneiderPRL12} To interpret
our data, we use recently developed theory of the graphite band
structure subject to a tilted magnetic field, which predicts lifting
of the twofold degeneracy at the $H$ point.\cite{Gonch_2012} This
degeneracy, taking origin in the 3D character of graphite (four
atoms in a unit cell instead of two for graphene),  is an
additional one to the valley and spin degeneracies in graphene.

\section{Experiment}

Thin graphite specimens for our magneto-transmission study have been
prepared by exfoliation. A thin layer of
bulk graphite, with an average thickness around $\approx$100~nm, was
located on the scotch tape used for exfoliation, which has several
relatively wide spectral windows with a sufficiently high optical
transmission. A high-quality natural graphite crystal has been
chosen for exfoliation, since it provides equivalent but better
pronounced magneto-optical response as compared to, \textit{e.g.},
highly-oriented pyrolytic graphite.\cite{OrlitaJPCM08} The
magneto-transmission spectra were measured for the magnetic field
inclined with respect to the $c$-axis of graphite by selected angles
of $\varphi$, \textsl{i.e.}, in the perpendicular
($\varphi=0^{\circ}$) and several tilted-field configurations.

To measure the transmission spectra in the spectral range 100-800~meV,
the non-polarized radiation of a globar was analyzed by a Fourier
transform spectrometer and guided to the sample by  light-pipe
optics. The sample was placed in a cryostat at temperature of 2~K
located inside superconducting and resistive coils, which reach magnetic
fields up to 13~T and 28~T, respectively. The transmitted signal was detected
by the composite Si bolometer. All spectra presented in this study have been
normalized by the zero-field transmission.

\section{Results}

\begin{figure}[t]
\begin{center}
\includegraphics[width=.85\linewidth]{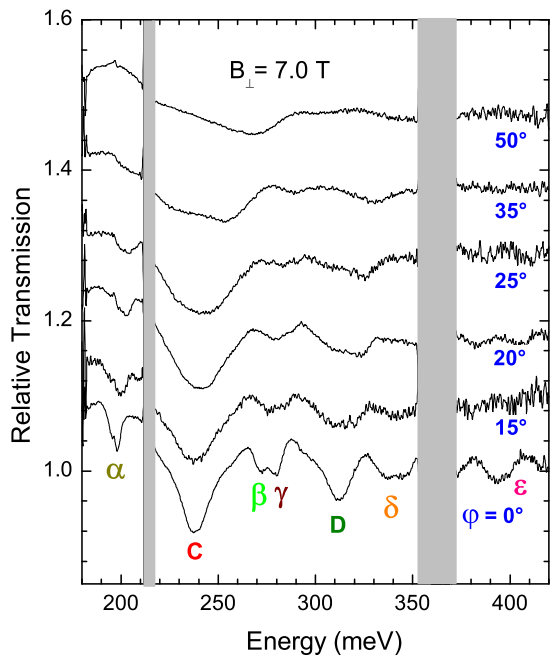}
\end{center}
\caption{\label{Fig2} Relative magneto-transmission spectra for tilting
  angles
  $\varphi=0^{\circ},15^{\circ},20^{\circ},25^{\circ},35^{\circ}$ and
  50$^{\circ}$ taken with the same perpendicular component of the
  field $B_\perp=7$~T. The broadening of the lines with increasing
  angle, \textit{i.e.}, with the increasing in-plane component of the
  field $B_\|$ is well visible, e.g., on the C line, the width
  of which increases roughly linearly with $B_\|$. A sample different
  from Fig.~\ref{Fig1} has been used (with higher density of
  crystallites and their average thickness). For clarity, successive spectra 
  are shifted vertically by 0.1.}
\end{figure}

The magneto-transmission spectra measured at three different angles
between the $c$-axis of graphite and magnetic field, $\varphi=0^{\circ},15^{\circ}$
and 30$^{\circ}$, are presented in Figs.~\ref{Fig1}a-c, respectively.
To facilitate identification of individual absorption lines, the
transmission curves are depicted as function of the photon energy normalized by
the factor of $\sqrt{B_\perp}$, which is typical of
LLs in a system of ideal 2D massless Dirac fermions. Plotted this way, we can
 identify graphene-like signal originating at the $H$ point and
easily follow its deviation from the $\sqrt{B_\perp}$ dependence
induced by the in-plane component of the magnetic field $B_\|$. An additional
set of data is presented in Fig.~\ref{Fig2}, where the
magneto-transmission spectra for several tilting angles $\varphi$
are presented, all measured at a fixed perpendicular component
of the field $B_\perp=7$~T.

The observed absorption lines have been marked consistently with
the notation introduced earlier.\cite{SadowskiPRL06,OrlitaPRL08}
The transitions denoted by Roman letters have their direct counterpart
in the response of graphene,\cite{SadowskiPRL06} while the ``Greek''
lines are characteristic of graphite.\cite{OrlitaPRL08} They are, in principle,
dipole-forbidden in a pure 2D system of Dirac fermions, nevertheless,
they can be consistently explained with the standard dipole selection
rule ($n\rightarrow n\pm1$) when the twofold degeneracy of LLs at the
$H$ points of bulk graphite is properly considered. Let us note that
the widths of absorption lines reflect not only the naturally present
disorder in the  graphite crystal, but they are also partially given by the
profile of the individual Landau subbands in the vicinity of the $H$
point.\cite{OrlitaSSC09}

\begin{figure}[t]
\begin{center}
\includegraphics[width=.85\linewidth]{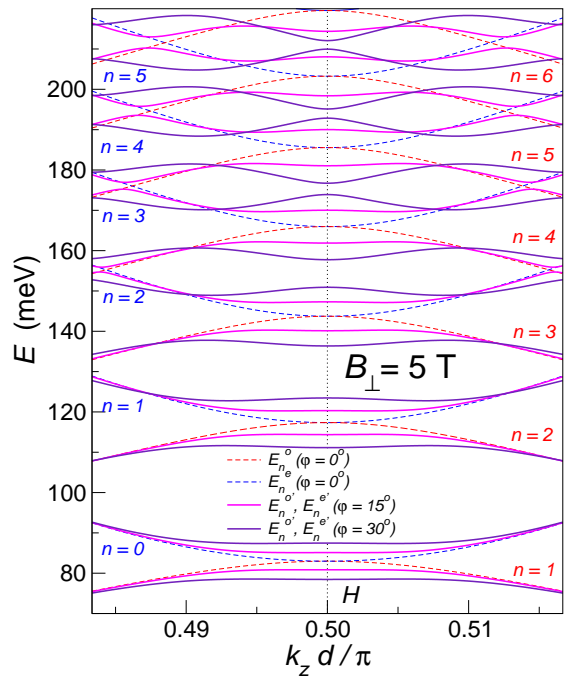}
\end{center}
\caption{\label{Fig3} Landau subbands in the vicinity of
the $H$ points, $k_zd=\pi/2$, at $B_\perp=$~5~T. The solid and dashed lines
describe the $k_z$ dependence of energies $E_n^e$ and $E_n^o$ in even and odd graphene
sheets at $\varphi=0^{\circ}, 15^{\circ}$ and 30$^{\circ}$.}
\end{figure}

The transmission spectra measured at $\varphi=0^{\circ}$, presented in
Fig.~\ref{Fig1}a, are fully analogous to previous
experiments\cite{OrlitaPRL08,OrlitaPRL09,ChuangPRB09} and the color
vertical lines mark transmission minima proportional to
$\sqrt{B_\perp}$. All such transitions originate at the $H$ point. The
transmission minima are more pronounced at lower energies and their
width increases with the increasing magnetic field. Interestingly, the
number of observed transitions remains nearly constant with
$B_\perp$. This behavior reflects the specific energy dependence of
the relaxation rate (\textit{i.e.}, broadening of lines), $\tau^{-1}(E)\propto|E|$,
which maps the (linear in energy) density of states around the $H$ point.
An analogous effect has been recently observed also in graphene
specimens.\cite{OrlitaPRL11}

At a non-zero tilting angle $\varphi$, the observed
magneto-transmission spectra significantly deviate from expectations
for a purely 2D system, which is in the case of orbital effects only sensitive
to the perpendicular component of the field. At $\varphi$ as low as
15$^{\circ}$, see Fig.~\ref{Fig1}b, the transitions denoted by Roman
letters change the shape and broaden, while the ``Greek'' lines become
significantly weaker.  For $\varphi>30^{\circ}$, the ``Greek'' lines
completely disappear from spectra and also the Roman lines are much less
pronounced as a result of a significant broadening. Alternatively, we
can follow these effects in Fig.~\ref{Fig2}, where the
magneto-transmission spectra are plotted at several angles
with the perpendicular component of the field kept constant, $B_\perp=7$~T.
The effects induced by the in-plane component of the field
$B_\|$ are well illustrated, \textit{e.g.}, on the C line.
With increasing tilting angle, this line does not only significantly
broaden, but also gains a complex structure -- a strong asymmetry is developed
and the line becomes nearly split into two components for higher angles
($\varphi>25^{^\circ}$).

\section{Discussion}

To interpret the broadening of absorption lines with $B_{||}$, we will
consider the electronic band structure at the $H$ point of graphite in detail.
In particular, we will focus on the $B_{||}$-induced lifting of the twofold degeneracy, which
in graphite complements the spin and valley degeneracies in graphene, and
follow the theory recently developed by Goncharuk and Smr\v{c}ka.\cite{Gonch_2012}
The additional twofold degeneracy may be interpreted as a direct
consequence of the effectively vanishing interlayer interaction for
a charge carrier with the momentum $k_z d=\pi/2$, where $d$ is the interlayer distance.
The reason is that the neighboring graphene sheets are rotated by 30$^{\circ}$. If the field dependence
of the energy in even layers is $E_n^e \propto\pm\sqrt{B_\perp n}$ then the energy in odd
layers reads $E_n^o \propto\pm\sqrt{B_\perp(n+1)}$, where $n =
0,1,\cdots$ is the index of the LLs.  Two states $|n+1\rangle^e$ and
$|n\rangle^o$ belonging to degenerated eigenenergies $E_{n+1}^e
=E_n^o$ are orthogonal and, therefore, the corresponding interlayer
hopping integral is equal to zero for $B_\|=0$. In
tilted fields the in-plane field component $B_{||}$ shifts the mutual
position of centers of orbits $|n+1\rangle^e$ and $|n\rangle^o$ in
real space by $y_0=d\,B_{||}/B_\perp$. The orbits are no longer
exactly orthogonal and the interlayer interaction does not completely
vanish.  In the lowest order of the perturbation theory we get instead
of $E_n^o$ and $E_{n+1}^e$ four energies $E_{n}^{o',e'} = \pm
T_{n,n+1}\pm [(E_n^o)^2 +T_{n,n+1}^2]^{1/2}$ where $T_{n,n+1} =
\gamma_1 d\, \tan(\varphi)\, \sqrt{B_\perp |e|(n+1)/(2\hbar)}$.\cite{Gonch_2012}
Let us note that another theory presented in
Ref.~\onlinecite{PershogubaPRB10} is devoted to the case of magnetic
field applied strictly parallel to the  sheets
of the graphene bilayer and multilayers including graphite. It is
suggested that
the obtained energy spectrum can be verified experimentally
using electron tunneling or optical spectroscopy.

Obviously, the first order perturbation theory employed in Ref.~\onlinecite{Gonch_2012},
which involves only two states with the same energy exactly at $k_zd=\pi/2$, is not
the best approximation and overestimates the splitting. It is acceptable
only for rather small tilting angles and LLs with limited $n$. For this
reason, we have calculated the eigenenergies numerically employing a
larger basis, which allows us to calculate also the $k_z$ dependence in
the vicinity of the $H$ points. The results are shown in Fig.~\ref{Fig3}.
Dashed lines describe the $k_z$ dependence of energies $E_n^e$ and $E_n^o$
in even and odd graphene sheets at $\varphi=0^{\circ}$. Solid lines describe the
energies $E_{n}^{e'}$ and $E_n^{o'}$, both at $\varphi=15^{\circ}$
and $30^{\circ}$, resolved by different colors. The
corresponding eigenstates mix the wave functions from
both even and odd layers, nevertheless, the same notation is kept
to emphasize to what energies $E_{n}^{e'}$ and $E_n^{o'}$ are
reduced for $\varphi \rightarrow 0$.

Our numerical calculations reveal the importance of
the states in the vicinity of the $H$ point. Two side extrema
in the $k_z$ dependence of energy subbands appear
and the curvature of the $k_z$ curves is reversed at the $H$
point. Each new local extremum developed along the $k_z$
dependence of the energy subbands adds a new van Hove singularity
to the (joint) density of states. The energy gap opened between
subbands appears slightly away of the $H$ point and it decreases with the
increasing LL index $n$ ($n\geq2$). Unlike the splitting
directly at the $H$ point, this energy gap of is no longer proportional
to $\sqrt{B_\perp}$.

At $\varphi=0$, each absorption line consists of up to four
degenerate transitions. To be more specific, we have two transitions
for each ``Greek'' line and also for the B line, the rest of lines
denoted by Roman letters include four degenerate transitions, with corresponding
four (degenerate) van Hove singularities in the joint density of states, for
details see Refs.~\onlinecite{OrlitaPRL08,OrlitaSSC09}. At $\varphi=0$, positions
of these singularities are represented in Fig.~\ref{Fig1}a by the vertical color lines.
The situation becomes more complex at $\varphi>0$, when the degeneracy of
these four Van Hove singularities is lifted. In addition, other singularities,
presumably weaker, develop in the vicinity of $H$ point. To illustrate the strength of effects
induced by the in-plane magnetic field, we plot, in Figs.~\ref{Fig1}b,c,
positions four main singularities in the limits of validity of the first
order approximation -- those which originate directly
at the $H$ point. Two of them remain almost degenerate.
On the other hand, for the sake of clarity, we do not mark positions
of additional singularities developed due to anti-crossing of Landau subbands
further from the $H$ point. To justify this, we note that such singularities in
principle exist even at $B_\|=0$ due to the trigonal warping term
$\gamma_3$,\cite{NakaoJPSJ76} which is neglected in the simple effective monolayer and
bilayer model. Nevertheless, they have not been observed in the experiment
in perpendicular fields.\cite{OrlitaPRL08}

To sum up, the in-plane magnetic field profoundly modifies the profile of Landau subbands
at the $H$ point of bulk graphite. A series of minigaps is created already within
the first order of the perturbation theory, which is directly reflected by newly
developed van Hove singularities in the joint density of states.
Experimentally, this leads to the splitting of the observed dipole-allowed transitions
with the increasing angle $\varphi$, or at least, if the disorder effects are realistically
involved, to a significant broadening of these transitions. This splitting/broadening increases
at the fixed $B_\perp$ roughly linearly with $B_\|$, or equivalently, approximatively scales
as $\sqrt{B_\perp}$ when the tilting angle $\varphi$ is kept constant.
Such behavior is consistent with the experimental data plotted in Fig.~\ref{Fig2} and in
Figs.~\ref{Fig1}a-c, respectively.

\section{Conclusions}

The electronic band structure at the $H$ point of bulk graphite has been
studied using the infrared transmission technique in magnetic fields tilted
with respect to the $c$-axis of this material. While for the magnetic field applied
along this axis, the magneto-optical responses of graphite (due to the $H$ point) and
graphene (due to the $K$ point) closely resemble each other, pronounced
deviations clearly appear with the increasing tilting angle. The 3D nature of the electronic
band structure of bulk graphite is thus revealed at significantly lower
tilting angles as compared to SdH and dHvA measurements.\cite{SchneiderPRB10,SchneiderPRL12}

\begin{acknowledgments}
We thank to M. Potemski for valuable discussions.
The support of the European Science Foundation EPIGRAT project (GRA/10/E006), GACR
No.~P204/10/1020, programme ``Transnational access'' contract No.~228043-EuroMagNET II-Integrated Activities,
AVCR research program AVOZ10100521, the Academy of Sciences of the Czech
Republic project KAN400100652 and the fund No.~SVV-2012-265306 via the Charles University
in Prague are acknowledged.
\end{acknowledgments}


\begin{thebibliography}{48}%
\makeatletter
\providecommand \@ifxundefined [1]{%
 \@ifx{#1\undefined}
}%
\providecommand \@ifnum [1]{%
 \ifnum #1\expandafter \@firstoftwo
 \else \expandafter \@secondoftwo
 \fi
}%
\providecommand \@ifx [1]{%
 \ifx #1\expandafter \@firstoftwo
 \else \expandafter \@secondoftwo
 \fi
}%
\providecommand \natexlab [1]{#1}%
\providecommand \enquote  [1]{``#1''}%
\providecommand \bibnamefont  [1]{#1}%
\providecommand \bibfnamefont [1]{#1}%
\providecommand \citenamefont [1]{#1}%
\providecommand \href@noop [0]{\@secondoftwo}%
\providecommand \href [0]{\begingroup \@sanitize@url \@href}%
\providecommand \@href[1]{\@@startlink{#1}\@@href}%
\providecommand \@@href[1]{\endgroup#1\@@endlink}%
\providecommand \@sanitize@url [0]{\catcode `\\12\catcode `\$12\catcode
  `\&12\catcode `\#12\catcode `\^12\catcode `\_12\catcode `\%12\relax}%
\providecommand \@@startlink[1]{}%
\providecommand \@@endlink[0]{}%
\providecommand \url  [0]{\begingroup\@sanitize@url \@url }%
\providecommand \@url [1]{\endgroup\@href {#1}{\urlprefix }}%
\providecommand \urlprefix  [0]{URL }%
\providecommand \Eprint [0]{\href }%
\providecommand \doibase [0]{http://dx.doi.org/}%
\providecommand \selectlanguage [0]{\@gobble}%
\providecommand \bibinfo  [0]{\@secondoftwo}%
\providecommand \bibfield  [0]{\@secondoftwo}%
\providecommand \translation [1]{[#1]}%
\providecommand \BibitemOpen [0]{}%
\providecommand \bibitemStop [0]{}%
\providecommand \bibitemNoStop [0]{.\EOS\space}%
\providecommand \EOS [0]{\spacefactor3000\relax}%
\providecommand \BibitemShut  [1]{\csname bibitem#1\endcsname}%
\let\auto@bib@innerbib\@empty
\bibitem [{\citenamefont {Berger}\ \emph {et~al.}(2004)\citenamefont {Berger},
  \citenamefont {Song}, \citenamefont {Li}, \citenamefont {Li}, \citenamefont
  {Ogbazghi}, \citenamefont {Feng}, \citenamefont {Dai}, \citenamefont
  {Marchenkov}, \citenamefont {Conrad}, \citenamefont {First},\ and\
  \citenamefont {de~Heer}}]{BergerJPCB04}%
  \BibitemOpen
  \bibfield  {author} {\bibinfo {author} {\bibfnamefont {C.}~\bibnamefont
  {Berger}}, \bibinfo {author} {\bibfnamefont {Z.}~\bibnamefont {Song}},
  \bibinfo {author} {\bibfnamefont {T.}~\bibnamefont {Li}}, \bibinfo {author}
  {\bibfnamefont {X.}~\bibnamefont {Li}}, \bibinfo {author} {\bibfnamefont
  {A.~Y.}\ \bibnamefont {Ogbazghi}}, \bibinfo {author} {\bibfnamefont
  {R.}~\bibnamefont {Feng}}, \bibinfo {author} {\bibfnamefont {Z.}~\bibnamefont
  {Dai}}, \bibinfo {author} {\bibfnamefont {A.~N.}\ \bibnamefont {Marchenkov}},
  \bibinfo {author} {\bibfnamefont {E.~H.}\ \bibnamefont {Conrad}}, \bibinfo
  {author} {\bibfnamefont {P.~N.}\ \bibnamefont {First}}, \ and\ \bibinfo
  {author} {\bibfnamefont {W.~A.}\ \bibnamefont {de~Heer}},\ }\href@noop {}
  {\bibfield  {journal} {\bibinfo  {journal} {J. Phys. Chem. B}\ }\textbf
  {\bibinfo {volume} {108}},\ \bibinfo {pages} {19912} (\bibinfo {year}
  {2004})}\BibitemShut {NoStop}%
\bibitem [{\citenamefont {Novoselov}\ \emph {et~al.}(2004)\citenamefont
  {Novoselov}, \citenamefont {Geim}, \citenamefont {Morozov}, \citenamefont
  {Jiang}, \citenamefont {Katsnelson}, \citenamefont {Grigorieva},
  \citenamefont {Dubonos},\ and\ \citenamefont {Firsov}}]{NovoselovScience04}%
  \BibitemOpen
  \bibfield  {author} {\bibinfo {author} {\bibfnamefont {K.~S.}\ \bibnamefont
  {Novoselov}}, \bibinfo {author} {\bibfnamefont {A.~K.}\ \bibnamefont {Geim}},
  \bibinfo {author} {\bibfnamefont {S.}~\bibnamefont {Morozov}}, \bibinfo
  {author} {\bibfnamefont {D.}~\bibnamefont {Jiang}}, \bibinfo {author}
  {\bibfnamefont {M.~I.}\ \bibnamefont {Katsnelson}}, \bibinfo {author}
  {\bibfnamefont {I.}~\bibnamefont {Grigorieva}}, \bibinfo {author}
  {\bibfnamefont {S.}~\bibnamefont {Dubonos}}, \ and\ \bibinfo {author}
  {\bibfnamefont {A.~A.}\ \bibnamefont {Firsov}},\ }\href@noop {} {\bibfield
  {journal} {\bibinfo  {journal} {Science}\ }\textbf {\bibinfo {volume}
  {306}},\ \bibinfo {pages} {666} (\bibinfo {year} {2004})}\BibitemShut
  {NoStop}%
\bibitem [{\citenamefont {Novoselov}\ \emph {et~al.}(2005)\citenamefont
  {Novoselov}, \citenamefont {Geim}, \citenamefont {Morozov}, \citenamefont
  {Jiang}, \citenamefont {Katsnelson}, \citenamefont {Grigorieva},
  \citenamefont {Dubonos},\ and\ \citenamefont {Firsov}}]{geim05}%
  \BibitemOpen
  \bibfield  {author} {\bibinfo {author} {\bibfnamefont {K.~S.}\ \bibnamefont
  {Novoselov}}, \bibinfo {author} {\bibfnamefont {A.~K.}\ \bibnamefont {Geim}},
  \bibinfo {author} {\bibfnamefont {S.~V.}\ \bibnamefont {Morozov}}, \bibinfo
  {author} {\bibfnamefont {D.}~\bibnamefont {Jiang}}, \bibinfo {author}
  {\bibfnamefont {M.~I.}\ \bibnamefont {Katsnelson}}, \bibinfo {author}
  {\bibfnamefont {I.~V.}\ \bibnamefont {Grigorieva}}, \bibinfo {author}
  {\bibfnamefont {S.~V.}\ \bibnamefont {Dubonos}}, \ and\ \bibinfo {author}
  {\bibfnamefont {A.~A.}\ \bibnamefont {Firsov}},\ }\href@noop {} {\bibfield
  {journal} {\bibinfo  {journal} {Nature}\ }\textbf {\bibinfo {volume} {438}},\
  \bibinfo {pages} {197} (\bibinfo {year} {2005})}\BibitemShut {NoStop}%
\bibitem [{\citenamefont {Zhang}\ \emph {et~al.}(2005)\citenamefont {Zhang},
  \citenamefont {Tan}, \citenamefont {Stormer},\ and\ \citenamefont
  {Kim}}]{Kim_2005}%
  \BibitemOpen
  \bibfield  {author} {\bibinfo {author} {\bibfnamefont {Y.}~\bibnamefont
  {Zhang}}, \bibinfo {author} {\bibfnamefont {Y.-W.}\ \bibnamefont {Tan}},
  \bibinfo {author} {\bibfnamefont {H.~L.}\ \bibnamefont {Stormer}}, \ and\
  \bibinfo {author} {\bibfnamefont {P.}~\bibnamefont {Kim}},\ }\href@noop {}
  {\bibfield  {journal} {\bibinfo  {journal} {Nature}\ }\textbf {\bibinfo
  {volume} {438}},\ \bibinfo {pages} {201} (\bibinfo {year}
  {2005})}\BibitemShut {NoStop}%
\bibitem [{\citenamefont {Li}\ \emph {et~al.}(2006)\citenamefont {Li},
  \citenamefont {Tsai}, \citenamefont {Padilla}, \citenamefont {Dordevic},
  \citenamefont {Burch}, \citenamefont {Wang},\ and\ \citenamefont
  {Basov}}]{LiPRB06}%
  \BibitemOpen
  \bibfield  {author} {\bibinfo {author} {\bibfnamefont {Z.~Q.}\ \bibnamefont
  {Li}}, \bibinfo {author} {\bibfnamefont {S.-W.}\ \bibnamefont {Tsai}},
  \bibinfo {author} {\bibfnamefont {W.~J.}\ \bibnamefont {Padilla}}, \bibinfo
  {author} {\bibfnamefont {S.~V.}\ \bibnamefont {Dordevic}}, \bibinfo {author}
  {\bibfnamefont {K.~S.}\ \bibnamefont {Burch}}, \bibinfo {author}
  {\bibfnamefont {Y.~J.}\ \bibnamefont {Wang}}, \ and\ \bibinfo {author}
  {\bibfnamefont {D.~N.}\ \bibnamefont {Basov}},\ }\href@noop {} {\bibfield
  {journal} {\bibinfo  {journal} {Phys. Rev. B}\ }\textbf {\bibinfo {volume}
  {74}},\ \bibinfo {pages} {195404} (\bibinfo {year} {2006})}\BibitemShut
  {NoStop}%
\bibitem [{\citenamefont {Luk'yanchuk}\ and\ \citenamefont
  {Kopelevich}(2006)}]{LukyanchukPRL06}%
  \BibitemOpen
  \bibfield  {author} {\bibinfo {author} {\bibfnamefont {I.~A.}\ \bibnamefont
  {Luk'yanchuk}}\ and\ \bibinfo {author} {\bibfnamefont {Y.}~\bibnamefont
  {Kopelevich}},\ }\href {\doibase 10.1103/PhysRevLett.97.256801} {\bibfield
  {journal} {\bibinfo  {journal} {Phys. Rev. Lett.}\ }\textbf {\bibinfo
  {volume} {97}},\ \bibinfo {pages} {256801} (\bibinfo {year}
  {2006})}\BibitemShut {NoStop}%
\bibitem [{\citenamefont {Gonz\'{a}lez}\ \emph {et~al.}(2007)\citenamefont
  {Gonz\'{a}lez}, \citenamefont {Munoz}, , \citenamefont {Garc\'{\i}a},
  \citenamefont {Barzola-Quiquia}, \citenamefont {Spoddig}, \citenamefont
  {Schindler}, ,\ and\ \citenamefont {Esquinazi}}]{GonzalesPRL07}%
  \BibitemOpen
  \bibfield  {author} {\bibinfo {author} {\bibfnamefont {J.~C.}\ \bibnamefont
  {Gonz\'{a}lez}}, \bibinfo {author} {\bibfnamefont {M.}~\bibnamefont {Munoz}},
  , \bibinfo {author} {\bibfnamefont {N.}~\bibnamefont {Garc\'{\i}a}}, \bibinfo
  {author} {\bibfnamefont {J.}~\bibnamefont {Barzola-Quiquia}}, \bibinfo
  {author} {\bibfnamefont {D.}~\bibnamefont {Spoddig}}, \bibinfo {author}
  {\bibfnamefont {K.}~\bibnamefont {Schindler}}, , \ and\ \bibinfo {author}
  {\bibfnamefont {P.}~\bibnamefont {Esquinazi}},\ }\href {\doibase
  10.1103/PhysRevLett.99.216601} {\bibfield  {journal} {\bibinfo  {journal}
  {Phys. Rev. Lett.}\ }\textbf {\bibinfo {volume} {99}},\ \bibinfo {pages}
  {216601} (\bibinfo {year} {2007})}\BibitemShut {NoStop}%
\bibitem [{\citenamefont {Orlita}\ \emph
  {et~al.}(2008{\natexlab{a}})\citenamefont {Orlita}, \citenamefont {Faugeras},
  \citenamefont {Martinez}, \citenamefont {Maude}, \citenamefont {Sadowski},\
  and\ \citenamefont {Potemski}}]{OrlitaPRL08}%
  \BibitemOpen
  \bibfield  {author} {\bibinfo {author} {\bibfnamefont {M.}~\bibnamefont
  {Orlita}}, \bibinfo {author} {\bibfnamefont {C.}~\bibnamefont {Faugeras}},
  \bibinfo {author} {\bibfnamefont {G.}~\bibnamefont {Martinez}}, \bibinfo
  {author} {\bibfnamefont {D.~K.}\ \bibnamefont {Maude}}, \bibinfo {author}
  {\bibfnamefont {M.~L.}\ \bibnamefont {Sadowski}}, \ and\ \bibinfo {author}
  {\bibfnamefont {M.}~\bibnamefont {Potemski}},\ }\href@noop {} {\bibfield
  {journal} {\bibinfo  {journal} {Phys. Rev. Lett.}\ }\textbf {\bibinfo
  {volume} {100}},\ \bibinfo {pages} {136403} (\bibinfo {year}
  {2008}{\natexlab{a}})}\BibitemShut {NoStop}%
\bibitem [{\citenamefont {Orlita}\ \emph
  {et~al.}(2008{\natexlab{b}})\citenamefont {Orlita}, \citenamefont {Faugeras},
  \citenamefont {Martinez}, \citenamefont {Maude}, \citenamefont {Sadowski},
  \citenamefont {Schneider},\ and\ \citenamefont {Potemski}}]{OrlitaJPCM08}%
  \BibitemOpen
  \bibfield  {author} {\bibinfo {author} {\bibfnamefont {M.}~\bibnamefont
  {Orlita}}, \bibinfo {author} {\bibfnamefont {C.}~\bibnamefont {Faugeras}},
  \bibinfo {author} {\bibfnamefont {G.}~\bibnamefont {Martinez}}, \bibinfo
  {author} {\bibfnamefont {D.~K.}\ \bibnamefont {Maude}}, \bibinfo {author}
  {\bibfnamefont {M.~L.}\ \bibnamefont {Sadowski}}, \bibinfo {author}
  {\bibfnamefont {J.~M.}\ \bibnamefont {Schneider}}, \ and\ \bibinfo {author}
  {\bibfnamefont {M.}~\bibnamefont {Potemski}},\ }\href@noop {} {\bibfield
  {journal} {\bibinfo  {journal} {J. Phys.: Condens. Mat.}\ }\textbf {\bibinfo
  {volume} {20}},\ \bibinfo {pages} {454223} (\bibinfo {year}
  {2008}{\natexlab{b}})}\BibitemShut {NoStop}%
\bibitem [{\citenamefont {Orlita}\ \emph
  {et~al.}(2009{\natexlab{a}})\citenamefont {Orlita}, \citenamefont {Faugeras},
  \citenamefont {Schneider}, \citenamefont {Martinez}, \citenamefont {Maude},\
  and\ \citenamefont {Potemski}}]{OrlitaPRL09}%
  \BibitemOpen
  \bibfield  {author} {\bibinfo {author} {\bibfnamefont {M.}~\bibnamefont
  {Orlita}}, \bibinfo {author} {\bibfnamefont {C.}~\bibnamefont {Faugeras}},
  \bibinfo {author} {\bibfnamefont {J.~M.}\ \bibnamefont {Schneider}}, \bibinfo
  {author} {\bibfnamefont {G.}~\bibnamefont {Martinez}}, \bibinfo {author}
  {\bibfnamefont {D.~K.}\ \bibnamefont {Maude}}, \ and\ \bibinfo {author}
  {\bibfnamefont {M.}~\bibnamefont {Potemski}},\ }\href@noop {} {\bibfield
  {journal} {\bibinfo  {journal} {Phys. Rev. Lett.}\ }\textbf {\bibinfo
  {volume} {102}},\ \bibinfo {pages} {166401} (\bibinfo {year}
  {2009}{\natexlab{a}})}\BibitemShut {NoStop}%
\bibitem [{\citenamefont {Chuang}\ \emph {et~al.}(2009)\citenamefont {Chuang},
  \citenamefont {Baker},\ and\ \citenamefont {Nicholas}}]{ChuangPRB09}%
  \BibitemOpen
  \bibfield  {author} {\bibinfo {author} {\bibfnamefont {K.-C.}\ \bibnamefont
  {Chuang}}, \bibinfo {author} {\bibfnamefont {A.~M.~R.}\ \bibnamefont
  {Baker}}, \ and\ \bibinfo {author} {\bibfnamefont {R.~J.}\ \bibnamefont
  {Nicholas}},\ }\href {\doibase 10.1103/PhysRevB.80.161410} {\bibfield
  {journal} {\bibinfo  {journal} {Phys. Rev. B}\ }\textbf {\bibinfo {volume}
  {80}},\ \bibinfo {pages} {161410} (\bibinfo {year} {2009})}\BibitemShut
  {NoStop}%
\bibitem [{\citenamefont {Zhu}\ \emph {et~al.}(2010)\citenamefont {Zhu},
  \citenamefont {Yang}, \citenamefont {Fauqu\'{e}}, \citenamefont
  {Kopelevich},\ and\ \citenamefont {Behnia}}]{ZhuNP10}%
  \BibitemOpen
  \bibfield  {author} {\bibinfo {author} {\bibfnamefont {Z.}~\bibnamefont
  {Zhu}}, \bibinfo {author} {\bibfnamefont {H.}~\bibnamefont {Yang}}, \bibinfo
  {author} {\bibfnamefont {B.}~\bibnamefont {Fauqu\'{e}}}, \bibinfo {author}
  {\bibfnamefont {Y.}~\bibnamefont {Kopelevich}}, \ and\ \bibinfo {author}
  {\bibfnamefont {K.}~\bibnamefont {Behnia}},\ }\href@noop {} {\bibfield
  {journal} {\bibinfo  {journal} {Nature Phys.}\ }\textbf {\bibinfo {volume}
  {6}},\ \bibinfo {pages} {26} (\bibinfo {year} {2010})}\BibitemShut {NoStop}%
\bibitem [{\citenamefont {Ubrig}\ \emph {et~al.}(2011)\citenamefont {Ubrig},
  \citenamefont {Plochocka}, \citenamefont {Kossacki}, \citenamefont {Orlita},
  \citenamefont {Maude}, \citenamefont {Portugall},\ and\ \citenamefont
  {Rikken}}]{Ubrig_2011}%
  \BibitemOpen
  \bibfield  {author} {\bibinfo {author} {\bibfnamefont {N.}~\bibnamefont
  {Ubrig}}, \bibinfo {author} {\bibfnamefont {P.}~\bibnamefont {Plochocka}},
  \bibinfo {author} {\bibfnamefont {P.}~\bibnamefont {Kossacki}}, \bibinfo
  {author} {\bibfnamefont {M.}~\bibnamefont {Orlita}}, \bibinfo {author}
  {\bibfnamefont {D.~K.}\ \bibnamefont {Maude}}, \bibinfo {author}
  {\bibfnamefont {O.}~\bibnamefont {Portugall}}, \ and\ \bibinfo {author}
  {\bibfnamefont {G.~L. J.~A.}\ \bibnamefont {Rikken}},\ }\href@noop {}
  {\bibfield  {journal} {\bibinfo  {journal} {Phys. Rev. B}\ }\textbf {\bibinfo
  {volume} {83}},\ \bibinfo {pages} {073401} (\bibinfo {year}
  {2011})}\BibitemShut {NoStop}%
\bibitem [{\citenamefont {Tung}\ \emph {et~al.}(2011)\citenamefont {Tung},
  \citenamefont {Cadden-Zimansky}, \citenamefont {Qi}, \citenamefont {Jiang},\
  and\ \citenamefont {Smirnov}}]{Tung_2011}%
  \BibitemOpen
  \bibfield  {author} {\bibinfo {author} {\bibfnamefont {L.~C.}\ \bibnamefont
  {Tung}}, \bibinfo {author} {\bibfnamefont {P.}~\bibnamefont
  {Cadden-Zimansky}}, \bibinfo {author} {\bibfnamefont {J.}~\bibnamefont {Qi}},
  \bibinfo {author} {\bibfnamefont {Z.}~\bibnamefont {Jiang}}, \ and\ \bibinfo
  {author} {\bibfnamefont {D.}~\bibnamefont {Smirnov}},\ }\href@noop {}
  {\bibfield  {journal} {\bibinfo  {journal} {Phys. Rev. B}\ }\textbf {\bibinfo
  {volume} {84}},\ \bibinfo {pages} {153405} (\bibinfo {year}
  {2011})}\BibitemShut {NoStop}%
\bibitem [{\citenamefont {Kossacki}\ \emph {et~al.}(2011)\citenamefont
  {Kossacki}, \citenamefont {Faugeras}, \citenamefont {K\"uhne}, \citenamefont
  {Orlita}, \citenamefont {Nicolet}, \citenamefont {Schneider}, \citenamefont
  {Basko}, \citenamefont {Latyshev},\ and\ \citenamefont
  {Potemski}}]{KossackiPRB11}%
  \BibitemOpen
  \bibfield  {author} {\bibinfo {author} {\bibfnamefont {P.}~\bibnamefont
  {Kossacki}}, \bibinfo {author} {\bibfnamefont {C.}~\bibnamefont {Faugeras}},
  \bibinfo {author} {\bibfnamefont {M.}~\bibnamefont {K\"uhne}}, \bibinfo
  {author} {\bibfnamefont {M.}~\bibnamefont {Orlita}}, \bibinfo {author}
  {\bibfnamefont {A.~A.~L.}\ \bibnamefont {Nicolet}}, \bibinfo {author}
  {\bibfnamefont {J.~M.}\ \bibnamefont {Schneider}}, \bibinfo {author}
  {\bibfnamefont {D.~M.}\ \bibnamefont {Basko}}, \bibinfo {author}
  {\bibfnamefont {Y.~I.}\ \bibnamefont {Latyshev}}, \ and\ \bibinfo {author}
  {\bibfnamefont {M.}~\bibnamefont {Potemski}},\ }\href@noop {} {\bibfield
  {journal} {\bibinfo  {journal} {Phys. Rev. B}\ }\textbf {\bibinfo {volume}
  {84}},\ \bibinfo {pages} {235138} (\bibinfo {year} {2011})}\BibitemShut
  {NoStop}%
\bibitem [{\citenamefont {Kim}\ \emph {et~al.}(2012)\citenamefont {Kim},
  \citenamefont {Ma}, \citenamefont {Imambekov}, \citenamefont {Kalugin},
  \citenamefont {Lombardo}, \citenamefont {Ferrari}, \citenamefont {Kono},\
  and\ \citenamefont {Smirnov}}]{KimPRB12}%
  \BibitemOpen
  \bibfield  {author} {\bibinfo {author} {\bibfnamefont {Y.}~\bibnamefont
  {Kim}}, \bibinfo {author} {\bibfnamefont {Y.}~\bibnamefont {Ma}}, \bibinfo
  {author} {\bibfnamefont {A.}~\bibnamefont {Imambekov}}, \bibinfo {author}
  {\bibfnamefont {N.~G.}\ \bibnamefont {Kalugin}}, \bibinfo {author}
  {\bibfnamefont {A.}~\bibnamefont {Lombardo}}, \bibinfo {author}
  {\bibfnamefont {A.~C.}\ \bibnamefont {Ferrari}}, \bibinfo {author}
  {\bibfnamefont {J.}~\bibnamefont {Kono}}, \ and\ \bibinfo {author}
  {\bibfnamefont {D.}~\bibnamefont {Smirnov}},\ }\href {\doibase
  10.1103/PhysRevB.85.121403} {\bibfield  {journal} {\bibinfo  {journal} {Phys.
  Rev. B}\ }\textbf {\bibinfo {volume} {85}},\ \bibinfo {pages} {121403}
  (\bibinfo {year} {2012})}\BibitemShut {NoStop}%
\bibitem [{\citenamefont {Levallois}\ \emph {et~al.}(2012)\citenamefont
  {Levallois}, \citenamefont {Tran},\ and\ \citenamefont
  {Kuzmenko}}]{LevalloisSSC12}%
  \BibitemOpen
  \bibfield  {author} {\bibinfo {author} {\bibfnamefont {J.}~\bibnamefont
  {Levallois}}, \bibinfo {author} {\bibfnamefont {M.}~\bibnamefont {Tran}}, \
  and\ \bibinfo {author} {\bibfnamefont {A.~B.}\ \bibnamefont {Kuzmenko}},\
  }\href {\doibase 10.1016/j.ssc.2012.04.036} {\bibfield  {journal} {\bibinfo
  {journal} {Solid State Communications}\ }\textbf {\bibinfo {volume} {152}},\
  \bibinfo {pages} {1294} (\bibinfo {year} {2012})}\BibitemShut {NoStop}%
\bibitem [{\citenamefont {Orlita}\ \emph
  {et~al.}(2009{\natexlab{b}})\citenamefont {Orlita}, \citenamefont {Faugeras},
  \citenamefont {Martinez}, \citenamefont {Maude}, \citenamefont {Schneider},
  \citenamefont {Sprinkle}, \citenamefont {Berger}, \citenamefont {de~Heer},\
  and\ \citenamefont {Potemski}}]{OrlitaSSC09}%
  \BibitemOpen
  \bibfield  {author} {\bibinfo {author} {\bibfnamefont {M.}~\bibnamefont
  {Orlita}}, \bibinfo {author} {\bibfnamefont {C.}~\bibnamefont {Faugeras}},
  \bibinfo {author} {\bibfnamefont {G.}~\bibnamefont {Martinez}}, \bibinfo
  {author} {\bibfnamefont {D.~K.}\ \bibnamefont {Maude}}, \bibinfo {author}
  {\bibfnamefont {J.~M.}\ \bibnamefont {Schneider}}, \bibinfo {author}
  {\bibfnamefont {M.}~\bibnamefont {Sprinkle}}, \bibinfo {author}
  {\bibfnamefont {C.}~\bibnamefont {Berger}}, \bibinfo {author} {\bibfnamefont
  {W.~A.}\ \bibnamefont {de~Heer}}, \ and\ \bibinfo {author} {\bibfnamefont
  {M.}~\bibnamefont {Potemski}},\ }\href@noop {} {\bibfield  {journal}
  {\bibinfo  {journal} {Solid State Commun.}\ }\textbf {\bibinfo {volume}
  {149}},\ \bibinfo {pages} {1128} (\bibinfo {year}
  {2009}{\natexlab{b}})}\BibitemShut {NoStop}%
\bibitem [{\citenamefont {Sadowski}\ \emph {et~al.}(2006)\citenamefont
  {Sadowski}, \citenamefont {Martinez}, \citenamefont {Potemski}, \citenamefont
  {Berger},\ and\ \citenamefont {de~Heer}}]{SadowskiPRL06}%
  \BibitemOpen
  \bibfield  {author} {\bibinfo {author} {\bibfnamefont {M.~L.}\ \bibnamefont
  {Sadowski}}, \bibinfo {author} {\bibfnamefont {G.}~\bibnamefont {Martinez}},
  \bibinfo {author} {\bibfnamefont {M.}~\bibnamefont {Potemski}}, \bibinfo
  {author} {\bibfnamefont {C.}~\bibnamefont {Berger}}, \ and\ \bibinfo {author}
  {\bibfnamefont {W.~A.}\ \bibnamefont {de~Heer}},\ }\href@noop {} {\bibfield
  {journal} {\bibinfo  {journal} {Phys. Rev. Lett.}\ }\textbf {\bibinfo
  {volume} {97}},\ \bibinfo {pages} {266405} (\bibinfo {year}
  {2006})}\BibitemShut {NoStop}%
\bibitem [{\citenamefont {Gusynin}\ \emph {et~al.}(2007)\citenamefont
  {Gusynin}, \citenamefont {Sharapov},\ and\ \citenamefont
  {Carbotte}}]{GusyninPRL07}%
  \BibitemOpen
  \bibfield  {author} {\bibinfo {author} {\bibfnamefont {V.~P.}\ \bibnamefont
  {Gusynin}}, \bibinfo {author} {\bibfnamefont {S.~G.}\ \bibnamefont
  {Sharapov}}, \ and\ \bibinfo {author} {\bibfnamefont {J.~P.}\ \bibnamefont
  {Carbotte}},\ }\href@noop {} {\bibfield  {journal} {\bibinfo  {journal}
  {Phys. Rev. Lett.}\ }\textbf {\bibinfo {volume} {98}},\ \bibinfo {pages}
  {157402} (\bibinfo {year} {2007})}\BibitemShut {NoStop}%
\bibitem [{\citenamefont {Jiang}\ \emph {et~al.}(2007)\citenamefont {Jiang},
  \citenamefont {Henriksen}, \citenamefont {Tung}, \citenamefont {Wang},
  \citenamefont {Schwartz}, \citenamefont {Han}, \citenamefont {Kim},\ and\
  \citenamefont {Stormer}}]{JiangPRL07}%
  \BibitemOpen
  \bibfield  {author} {\bibinfo {author} {\bibfnamefont {Z.}~\bibnamefont
  {Jiang}}, \bibinfo {author} {\bibfnamefont {E.~A.}\ \bibnamefont
  {Henriksen}}, \bibinfo {author} {\bibfnamefont {L.~C.}\ \bibnamefont {Tung}},
  \bibinfo {author} {\bibfnamefont {Y.-J.}\ \bibnamefont {Wang}}, \bibinfo
  {author} {\bibfnamefont {M.~E.}\ \bibnamefont {Schwartz}}, \bibinfo {author}
  {\bibfnamefont {M.~Y.}\ \bibnamefont {Han}}, \bibinfo {author} {\bibfnamefont
  {P.}~\bibnamefont {Kim}}, \ and\ \bibinfo {author} {\bibfnamefont {H.~L.}\
  \bibnamefont {Stormer}},\ }\href@noop {} {\bibfield  {journal} {\bibinfo
  {journal} {Phys. Rev. Lett.}\ }\textbf {\bibinfo {volume} {98}},\ \bibinfo
  {pages} {197403} (\bibinfo {year} {2007})}\BibitemShut {NoStop}%
\bibitem [{\citenamefont {Deacon}\ \emph {et~al.}(2007)\citenamefont {Deacon},
  \citenamefont {Chuang}, \citenamefont {Nicholas}, \citenamefont {Novoselov},\
  and\ \citenamefont {Geim}}]{DeaconPRB07}%
  \BibitemOpen
  \bibfield  {author} {\bibinfo {author} {\bibfnamefont {R.~S.}\ \bibnamefont
  {Deacon}}, \bibinfo {author} {\bibfnamefont {K.-C.}\ \bibnamefont {Chuang}},
  \bibinfo {author} {\bibfnamefont {R.~J.}\ \bibnamefont {Nicholas}}, \bibinfo
  {author} {\bibfnamefont {K.~S.}\ \bibnamefont {Novoselov}}, \ and\ \bibinfo
  {author} {\bibfnamefont {A.~K.}\ \bibnamefont {Geim}},\ }\href@noop {}
  {\bibfield  {journal} {\bibinfo  {journal} {Phys. Rev. B}\ }\textbf {\bibinfo
  {volume} {76}},\ \bibinfo {pages} {081406R} (\bibinfo {year}
  {2007})}\BibitemShut {NoStop}%
\bibitem [{\citenamefont {Orlita}\ \emph
  {et~al.}(2008{\natexlab{c}})\citenamefont {Orlita}, \citenamefont {Faugeras},
  \citenamefont {Plochocka}, \citenamefont {Neugebauer}, \citenamefont
  {Martinez}, \citenamefont {Maude}, \citenamefont {Barra}, \citenamefont
  {Sprinkle}, \citenamefont {Berger}, \citenamefont {de~Heer},\ and\
  \citenamefont {Potemski}}]{OrlitaPRL08II}%
  \BibitemOpen
  \bibfield  {author} {\bibinfo {author} {\bibfnamefont {M.}~\bibnamefont
  {Orlita}}, \bibinfo {author} {\bibfnamefont {C.}~\bibnamefont {Faugeras}},
  \bibinfo {author} {\bibfnamefont {P.}~\bibnamefont {Plochocka}}, \bibinfo
  {author} {\bibfnamefont {P.}~\bibnamefont {Neugebauer}}, \bibinfo {author}
  {\bibfnamefont {G.}~\bibnamefont {Martinez}}, \bibinfo {author}
  {\bibfnamefont {D.~K.}\ \bibnamefont {Maude}}, \bibinfo {author}
  {\bibfnamefont {A.-L.}\ \bibnamefont {Barra}}, \bibinfo {author}
  {\bibfnamefont {M.}~\bibnamefont {Sprinkle}}, \bibinfo {author}
  {\bibfnamefont {C.}~\bibnamefont {Berger}}, \bibinfo {author} {\bibfnamefont
  {W.~A.}\ \bibnamefont {de~Heer}}, \ and\ \bibinfo {author} {\bibfnamefont
  {M.}~\bibnamefont {Potemski}},\ }\href@noop {} {\bibfield  {journal}
  {\bibinfo  {journal} {Phys. Rev. Lett.}\ }\textbf {\bibinfo {volume} {101}},\
  \bibinfo {pages} {267601} (\bibinfo {year} {2008}{\natexlab{c}})}\BibitemShut
  {NoStop}%
\bibitem [{\citenamefont {Crassee}\ \emph {et~al.}(2011)\citenamefont
  {Crassee}, \citenamefont {Levallois}, \citenamefont {Walter}, \citenamefont
  {Ostler}, \citenamefont {Bostwick}, \citenamefont {Rotenberg}, \citenamefont
  {Seyller}, \citenamefont {van~der Marel},\ and\ \citenamefont
  {Kuzmenko}}]{CrasseeNP11}%
  \BibitemOpen
  \bibfield  {author} {\bibinfo {author} {\bibfnamefont {I.}~\bibnamefont
  {Crassee}}, \bibinfo {author} {\bibfnamefont {J.}~\bibnamefont {Levallois}},
  \bibinfo {author} {\bibfnamefont {A.~L.}\ \bibnamefont {Walter}}, \bibinfo
  {author} {\bibfnamefont {M.}~\bibnamefont {Ostler}}, \bibinfo {author}
  {\bibfnamefont {A.}~\bibnamefont {Bostwick}}, \bibinfo {author}
  {\bibfnamefont {E.}~\bibnamefont {Rotenberg}}, \bibinfo {author}
  {\bibfnamefont {T.}~\bibnamefont {Seyller}}, \bibinfo {author} {\bibfnamefont
  {D.}~\bibnamefont {van~der Marel}}, \ and\ \bibinfo {author} {\bibfnamefont
  {A.~B.}\ \bibnamefont {Kuzmenko}},\ }\href@noop {} {\bibfield  {journal}
  {\bibinfo  {journal} {Nature Phys.}\ }\textbf {\bibinfo {volume} {7}},\
  \bibinfo {pages} {48} (\bibinfo {year} {2011})}\BibitemShut {NoStop}%
\bibitem [{\citenamefont {{ I. Crassee \textit{et al.}}}(2011)}]{CrasseePRB11}%
  \BibitemOpen
  \bibfield  {author} {\bibinfo {author} {\bibnamefont {{ I. Crassee \textit{et
  al.}}}},\ }\href@noop {} {\bibfield  {journal} {\bibinfo  {journal} {Phys.
  Rev. B}\ }\textbf {\bibinfo {volume} {84}},\ \bibinfo {pages} {035103}
  (\bibinfo {year} {2011})}\BibitemShut {NoStop}%
\bibitem [{\citenamefont {Booshehri}\ \emph {et~al.}(2012)\citenamefont
  {Booshehri}, \citenamefont {Mielke}, \citenamefont {Rickel}, \citenamefont
  {Crooker}, \citenamefont {Zhang}, \citenamefont {Ren}, \citenamefont
  {H\'aroz}, \citenamefont {Rustagi}, \citenamefont {Stanton}, \citenamefont
  {Jin}, \citenamefont {Sun}, \citenamefont {Yan}, \citenamefont {Tour},\ and\
  \citenamefont {Kono}}]{BooshehriPRB12}%
  \BibitemOpen
  \bibfield  {author} {\bibinfo {author} {\bibfnamefont {L.~G.}\ \bibnamefont
  {Booshehri}}, \bibinfo {author} {\bibfnamefont {C.~H.}\ \bibnamefont
  {Mielke}}, \bibinfo {author} {\bibfnamefont {D.~G.}\ \bibnamefont {Rickel}},
  \bibinfo {author} {\bibfnamefont {S.~A.}\ \bibnamefont {Crooker}}, \bibinfo
  {author} {\bibfnamefont {Q.}~\bibnamefont {Zhang}}, \bibinfo {author}
  {\bibfnamefont {L.}~\bibnamefont {Ren}}, \bibinfo {author} {\bibfnamefont
  {E.~H.}\ \bibnamefont {H\'aroz}}, \bibinfo {author} {\bibfnamefont
  {A.}~\bibnamefont {Rustagi}}, \bibinfo {author} {\bibfnamefont {C.~J.}\
  \bibnamefont {Stanton}}, \bibinfo {author} {\bibfnamefont {Z.}~\bibnamefont
  {Jin}}, \bibinfo {author} {\bibfnamefont {Z.}~\bibnamefont {Sun}}, \bibinfo
  {author} {\bibfnamefont {Z.}~\bibnamefont {Yan}}, \bibinfo {author}
  {\bibfnamefont {J.~M.}\ \bibnamefont {Tour}}, \ and\ \bibinfo {author}
  {\bibfnamefont {J.}~\bibnamefont {Kono}},\ }\href {\doibase
  10.1103/PhysRevB.85.205407} {\bibfield  {journal} {\bibinfo  {journal} {Phys.
  Rev. B}\ }\textbf {\bibinfo {volume} {85}},\ \bibinfo {pages} {205407}
  (\bibinfo {year} {2012})}\BibitemShut {NoStop}%
\bibitem [{\citenamefont {Toy}\ \emph {et~al.}(1977)\citenamefont {Toy},
  \citenamefont {Dressehaus},\ and\ \citenamefont {Dresselhaus}}]{ToyPRB77}%
  \BibitemOpen
  \bibfield  {author} {\bibinfo {author} {\bibfnamefont {T.~W.~W.}\
  \bibnamefont {Toy}}, \bibinfo {author} {\bibfnamefont {M.~S.}\ \bibnamefont
  {Dressehaus}}, \ and\ \bibinfo {author} {\bibfnamefont {G.}~\bibnamefont
  {Dresselhaus}},\ }\href@noop {} {\bibfield  {journal} {\bibinfo  {journal}
  {Phys. Rev. B}\ }\textbf {\bibinfo {volume} {15}},\ \bibinfo {pages} {4077}
  (\bibinfo {year} {1977})}\BibitemShut {NoStop}%
\bibitem [{\citenamefont {Galt}\ \emph {et~al.}(1956)\citenamefont {Galt},
  \citenamefont {Yager},\ and\ \citenamefont {Dail}}]{GaltPR56}%
  \BibitemOpen
  \bibfield  {author} {\bibinfo {author} {\bibfnamefont {J.~K.}\ \bibnamefont
  {Galt}}, \bibinfo {author} {\bibfnamefont {W.~A.}\ \bibnamefont {Yager}}, \
  and\ \bibinfo {author} {\bibfnamefont {H.~W.}\ \bibnamefont {Dail}},\
  }\href@noop {} {\bibfield  {journal} {\bibinfo  {journal} {Phys. Rev.}\
  }\textbf {\bibinfo {volume} {103}},\ \bibinfo {pages} {1586} (\bibinfo {year}
  {1956})}\BibitemShut {NoStop}%
\bibitem [{\citenamefont {Schroeder}\ \emph {et~al.}(1969)\citenamefont
  {Schroeder}, \citenamefont {Dresselhaus},\ and\ \citenamefont
  {Javan}}]{SchroederPRL68}%
  \BibitemOpen
  \bibfield  {author} {\bibinfo {author} {\bibfnamefont {P.~R.}\ \bibnamefont
  {Schroeder}}, \bibinfo {author} {\bibfnamefont {M.~S.}\ \bibnamefont
  {Dresselhaus}}, \ and\ \bibinfo {author} {\bibfnamefont {A.}~\bibnamefont
  {Javan}},\ }\href@noop {} {\bibfield  {journal} {\bibinfo  {journal} {Phys.
  Rev. Lett.}\ }\textbf {\bibinfo {volume} {20}},\ \bibinfo {pages} {1292}
  (\bibinfo {year} {1969})}\BibitemShut {NoStop}%
\bibitem [{\citenamefont {Slonczewski}\ and\ \citenamefont
  {Weiss}(1958)}]{S_W_1958}%
  \BibitemOpen
  \bibfield  {author} {\bibinfo {author} {\bibfnamefont {J.~C.}\ \bibnamefont
  {Slonczewski}}\ and\ \bibinfo {author} {\bibfnamefont {P.~R.}\ \bibnamefont
  {Weiss}},\ }\href@noop {} {\bibfield  {journal} {\bibinfo  {journal} {Phys.
  Rev.}\ }\textbf {\bibinfo {volume} {109}},\ \bibinfo {pages} {272} (\bibinfo
  {year} {1958})}\BibitemShut {NoStop}%
\bibitem [{\citenamefont {McClure}(1960)}]{McClure_1960}%
  \BibitemOpen
  \bibfield  {author} {\bibinfo {author} {\bibfnamefont {J.~W.}\ \bibnamefont
  {McClure}},\ }\href@noop {} {\bibfield  {journal} {\bibinfo  {journal} {Phys.
  Rev.}\ }\textbf {\bibinfo {volume} {119}},\ \bibinfo {pages} {606} (\bibinfo
  {year} {1960})}\BibitemShut {NoStop}%
\bibitem [{\citenamefont {Ho}\ \emph {et~al.}(2011{\natexlab{a}})\citenamefont
  {Ho}, \citenamefont {Chiu}, \citenamefont {Su},\ and\ \citenamefont
  {Lin}}]{HoAPL11}%
  \BibitemOpen
  \bibfield  {author} {\bibinfo {author} {\bibfnamefont {Y.-H.}\ \bibnamefont
  {Ho}}, \bibinfo {author} {\bibfnamefont {Y.-H.}\ \bibnamefont {Chiu}},
  \bibinfo {author} {\bibfnamefont {W.-P.}\ \bibnamefont {Su}}, \ and\ \bibinfo
  {author} {\bibfnamefont {M.-F.}\ \bibnamefont {Lin}},\ }\href@noop {}
  {\bibfield  {journal} {\bibinfo  {journal} {Applied Physics Letters}\
  }\textbf {\bibinfo {volume} {99}},\ \bibinfo {eid} {011914} (\bibinfo {year}
  {2011}{\natexlab{a}})}\BibitemShut {NoStop}%
\bibitem [{\citenamefont {Ho}\ \emph {et~al.}(2011{\natexlab{b}})\citenamefont
  {Ho}, \citenamefont {Wang}, \citenamefont {Chiu}, \citenamefont {Lin},\ and\
  \citenamefont {Su}}]{HoPRB11}%
  \BibitemOpen
  \bibfield  {author} {\bibinfo {author} {\bibfnamefont {Y.~H.}\ \bibnamefont
  {Ho}}, \bibinfo {author} {\bibfnamefont {J.}~\bibnamefont {Wang}}, \bibinfo
  {author} {\bibfnamefont {Y.~H.}\ \bibnamefont {Chiu}}, \bibinfo {author}
  {\bibfnamefont {M.~F.}\ \bibnamefont {Lin}}, \ and\ \bibinfo {author}
  {\bibfnamefont {W.~P.}\ \bibnamefont {Su}},\ }\href@noop {} {\bibfield
  {journal} {\bibinfo  {journal} {Phys. Rev. B}\ }\textbf {\bibinfo {volume}
  {83}},\ \bibinfo {pages} {121201} (\bibinfo {year}
  {2011}{\natexlab{b}})}\BibitemShut {NoStop}%
\bibitem [{\citenamefont {Brandt}\ \emph {et~al.}(1988)\citenamefont {Brandt},
  \citenamefont {Chudinov},\ and\ \citenamefont {Ponomarev}}]{Brandt88}%
  \BibitemOpen
  \bibfield  {author} {\bibinfo {author} {\bibfnamefont {N.~B.}\ \bibnamefont
  {Brandt}}, \bibinfo {author} {\bibfnamefont {S.~M.}\ \bibnamefont
  {Chudinov}}, \ and\ \bibinfo {author} {\bibfnamefont {Y.~G.}\ \bibnamefont
  {Ponomarev}},\ }\href@noop {} {\emph {\bibinfo {title} {Semimetals 1:
  Graphite and its Compounds}}},\ \bibinfo {series} {Modern Problems in
  Condensed Matter Sciences}, Vol.\ \bibinfo {volume} {20.1}\ (\bibinfo
  {publisher} {North-Holland, Amsterdam},\ \bibinfo {year} {1988})\BibitemShut
  {NoStop}%
\bibitem [{\citenamefont {Gr\"{u}neis}\ \emph {et~al.}(2008)\citenamefont
  {Gr\"{u}neis}, \citenamefont {Attaccalite}, \citenamefont {Wirtz},
  \citenamefont {Shiozawa}, \citenamefont {Saito}, \citenamefont {Pichler},\
  and\ \citenamefont {Rubio}}]{Gruneis_2008}%
  \BibitemOpen
  \bibfield  {author} {\bibinfo {author} {\bibfnamefont {A.}~\bibnamefont
  {Gr\"{u}neis}}, \bibinfo {author} {\bibfnamefont {C.}~\bibnamefont
  {Attaccalite}}, \bibinfo {author} {\bibfnamefont {L.}~\bibnamefont {Wirtz}},
  \bibinfo {author} {\bibfnamefont {H.}~\bibnamefont {Shiozawa}}, \bibinfo
  {author} {\bibfnamefont {R.}~\bibnamefont {Saito}}, \bibinfo {author}
  {\bibfnamefont {T.}~\bibnamefont {Pichler}}, \ and\ \bibinfo {author}
  {\bibfnamefont {A.}~\bibnamefont {Rubio}},\ }\href@noop {} {\bibfield
  {journal} {\bibinfo  {journal} {Phys. Rev. B}\ }\textbf {\bibinfo {volume}
  {78}},\ \bibinfo {pages} {205425} (\bibinfo {year} {2008})}\BibitemShut
  {NoStop}%
\bibitem [{\citenamefont {Woollam}(1970)}]{WoollamPRL70}%
  \BibitemOpen
  \bibfield  {author} {\bibinfo {author} {\bibfnamefont {J.~A.}\ \bibnamefont
  {Woollam}},\ }\href {\doibase 10.1103/PhysRevLett.25.810} {\bibfield
  {journal} {\bibinfo  {journal} {Phys. Rev. Lett.}\ }\textbf {\bibinfo
  {volume} {25}},\ \bibinfo {pages} {810} (\bibinfo {year} {1970})}\BibitemShut
  {NoStop}%
\bibitem [{\citenamefont {Luk'yanchuk}\ and\ \citenamefont
  {Kopelevich}(2004)}]{LukyanchukPRL04}%
  \BibitemOpen
  \bibfield  {author} {\bibinfo {author} {\bibfnamefont {I.~A.}\ \bibnamefont
  {Luk'yanchuk}}\ and\ \bibinfo {author} {\bibfnamefont {Y.}~\bibnamefont
  {Kopelevich}},\ }\href {\doibase 10.1103/PhysRevLett.93.166402} {\bibfield
  {journal} {\bibinfo  {journal} {Phys. Rev. Lett.}\ }\textbf {\bibinfo
  {volume} {93}},\ \bibinfo {pages} {166402} (\bibinfo {year}
  {2004})}\BibitemShut {NoStop}%
\bibitem [{\citenamefont {Schneider}\ \emph {et~al.}(2010)\citenamefont
  {Schneider}, \citenamefont {Goncharuk}, \citenamefont {Va\v{s}ek},
  \citenamefont {Svoboda}, \citenamefont {V\'{y}born\'{y}}, \citenamefont
  {Smr\v{c}ka}, \citenamefont {Orlita}, \citenamefont {Potemski},\ and\
  \citenamefont {Maude}}]{SchneiderPRB10}%
  \BibitemOpen
  \bibfield  {author} {\bibinfo {author} {\bibfnamefont {J.~M.}\ \bibnamefont
  {Schneider}}, \bibinfo {author} {\bibfnamefont {N.~A.}\ \bibnamefont
  {Goncharuk}}, \bibinfo {author} {\bibfnamefont {P.}~\bibnamefont
  {Va\v{s}ek}}, \bibinfo {author} {\bibfnamefont {P.}~\bibnamefont {Svoboda}},
  \bibinfo {author} {\bibfnamefont {Z.}~\bibnamefont {V\'{y}born\'{y}}},
  \bibinfo {author} {\bibfnamefont {L.}~\bibnamefont {Smr\v{c}ka}}, \bibinfo
  {author} {\bibfnamefont {M.}~\bibnamefont {Orlita}}, \bibinfo {author}
  {\bibfnamefont {M.}~\bibnamefont {Potemski}}, \ and\ \bibinfo {author}
  {\bibfnamefont {D.~K.}\ \bibnamefont {Maude}},\ }\href@noop {} {\bibfield
  {journal} {\bibinfo  {journal} {Phys. Rev. B}\ }\textbf {\bibinfo {volume}
  {81}},\ \bibinfo {pages} {195204} (\bibinfo {year} {2010})}\BibitemShut
  {NoStop}%
\bibitem [{\citenamefont {Williamson}\ \emph {et~al.}(1965)\citenamefont
  {Williamson}, \citenamefont {Foner},\ and\ \citenamefont
  {Dresselhaus}}]{WilliamsonPR65}%
  \BibitemOpen
  \bibfield  {author} {\bibinfo {author} {\bibfnamefont {S.~J.}\ \bibnamefont
  {Williamson}}, \bibinfo {author} {\bibfnamefont {S.}~\bibnamefont {Foner}}, \
  and\ \bibinfo {author} {\bibfnamefont {M.~S.}\ \bibnamefont {Dresselhaus}},\
  }\href {\doibase 10.1103/PhysRev.140.A1429} {\bibfield  {journal} {\bibinfo
  {journal} {Phys. Rev.}\ }\textbf {\bibinfo {volume} {140}},\ \bibinfo {pages}
  {A1429} (\bibinfo {year} {1965})}\BibitemShut {NoStop}%
\bibitem [{\citenamefont {Schneider}\ \emph {et~al.}(2012)\citenamefont
  {Schneider}, \citenamefont {Piot}, \citenamefont {Sheikin},\ and\
  \citenamefont {Maude}}]{SchneiderPRL12}%
  \BibitemOpen
  \bibfield  {author} {\bibinfo {author} {\bibfnamefont {J.~M.}\ \bibnamefont
  {Schneider}}, \bibinfo {author} {\bibfnamefont {B.~A.}\ \bibnamefont {Piot}},
  \bibinfo {author} {\bibfnamefont {I.}~\bibnamefont {Sheikin}}, \ and\
  \bibinfo {author} {\bibfnamefont {D.~K.}\ \bibnamefont {Maude}},\ }\href@noop
  {} {\bibfield  {journal} {\bibinfo  {journal} {Phys. Rev. Lett.}\ }\textbf
  {\bibinfo {volume} {108}},\ \bibinfo {pages} {117401} (\bibinfo {year}
  {2012})}\BibitemShut {NoStop}%
\bibitem [{\citenamefont {Doezema}\ \emph {et~al.}(1979)\citenamefont
  {Doezema}, \citenamefont {Datars}, \citenamefont {Schaber},\ and\
  \citenamefont {Van~Schyndel}}]{DoezemaPRB79}%
  \BibitemOpen
  \bibfield  {author} {\bibinfo {author} {\bibfnamefont {R.~E.}\ \bibnamefont
  {Doezema}}, \bibinfo {author} {\bibfnamefont {W.~R.}\ \bibnamefont {Datars}},
  \bibinfo {author} {\bibfnamefont {H.}~\bibnamefont {Schaber}}, \ and\
  \bibinfo {author} {\bibfnamefont {A.}~\bibnamefont {Van~Schyndel}},\
  }\href@noop {} {\bibfield  {journal} {\bibinfo  {journal} {Phys. Rev. B}\
  }\textbf {\bibinfo {volume} {19}},\ \bibinfo {pages} {4224} (\bibinfo {year}
  {1979})}\BibitemShut {NoStop}%
\bibitem [{\citenamefont {Orlita}\ \emph {et~al.}(2012)\citenamefont {Orlita},
  \citenamefont {Neugebauer}, \citenamefont {Faugeras}, \citenamefont {Barra},
  \citenamefont {Potemski}, \citenamefont {Pellegrino},\ and\ \citenamefont
  {Basko}}]{OrlitaPRL12}%
  \BibitemOpen
  \bibfield  {author} {\bibinfo {author} {\bibfnamefont {M.}~\bibnamefont
  {Orlita}}, \bibinfo {author} {\bibfnamefont {P.}~\bibnamefont {Neugebauer}},
  \bibinfo {author} {\bibfnamefont {C.}~\bibnamefont {Faugeras}}, \bibinfo
  {author} {\bibfnamefont {A.-L.}\ \bibnamefont {Barra}}, \bibinfo {author}
  {\bibfnamefont {M.}~\bibnamefont {Potemski}}, \bibinfo {author}
  {\bibfnamefont {F.~M.~D.}\ \bibnamefont {Pellegrino}}, \ and\ \bibinfo
  {author} {\bibfnamefont {D.~M.}\ \bibnamefont {Basko}},\ }\href {\doibase
  10.1103/PhysRevLett.108.017602} {\bibfield  {journal} {\bibinfo  {journal}
  {Phys. Rev. Lett.}\ }\textbf {\bibinfo {volume} {108}},\ \bibinfo {pages}
  {017602} (\bibinfo {year} {2012})}\BibitemShut {NoStop}%
\bibitem [{\citenamefont {Goncharuk}\ and\ \citenamefont
  {Smr\v{c}ka}(2012)}]{Gonch_2012}%
  \BibitemOpen
  \bibfield  {author} {\bibinfo {author} {\bibfnamefont {N.~A.}\ \bibnamefont
  {Goncharuk}}\ and\ \bibinfo {author} {\bibfnamefont {L.}~\bibnamefont
  {Smr\v{c}ka}},\ }\href@noop {} {\bibfield  {journal} {\bibinfo  {journal}
  {Journal of Physics: Condensed Matter}\ }\textbf {\bibinfo {volume} {24}},\
  \bibinfo {pages} {185503} (\bibinfo {year} {2012})}\BibitemShut {NoStop}%
\bibitem [{\citenamefont {Koshino}\ and\ \citenamefont
  {Ando}(2008)}]{Kos_2008}%
  \BibitemOpen
  \bibfield  {author} {\bibinfo {author} {\bibfnamefont {M.}~\bibnamefont
  {Koshino}}\ and\ \bibinfo {author} {\bibfnamefont {T.}~\bibnamefont {Ando}},\
  }\href@noop {} {\bibfield  {journal} {\bibinfo  {journal} {Phys. Rev. B}\
  }\textbf {\bibinfo {volume} {77}},\ \bibinfo {pages} {115313} (\bibinfo
  {year} {2008})}\BibitemShut {NoStop}%
\bibitem [{\citenamefont {Partoens}\ and\ \citenamefont
  {Peeters}(2007)}]{PartoensPRB07}%
  \BibitemOpen
  \bibfield  {author} {\bibinfo {author} {\bibfnamefont {B.}~\bibnamefont
  {Partoens}}\ and\ \bibinfo {author} {\bibfnamefont {F.~M.}\ \bibnamefont
  {Peeters}},\ }\href@noop {} {\bibfield  {journal} {\bibinfo  {journal} {Phys.
  Rev. B}\ }\textbf {\bibinfo {volume} {75}},\ \bibinfo {pages} {193402}
  (\bibinfo {year} {2007})}\BibitemShut {NoStop}%
\bibitem [{\citenamefont {Orlita}\ \emph {et~al.}(2011)\citenamefont {Orlita},
  \citenamefont {Faugeras}, \citenamefont {Grill}, \citenamefont {Wysmolek},
  \citenamefont {Strupinski}, \citenamefont {Berger}, \citenamefont {de~Heer},
  \citenamefont {Martinez},\ and\ \citenamefont {Potemski}}]{OrlitaPRL11}%
  \BibitemOpen
  \bibfield  {author} {\bibinfo {author} {\bibfnamefont {M.}~\bibnamefont
  {Orlita}}, \bibinfo {author} {\bibfnamefont {C.}~\bibnamefont {Faugeras}},
  \bibinfo {author} {\bibfnamefont {R.}~\bibnamefont {Grill}}, \bibinfo
  {author} {\bibfnamefont {A.}~\bibnamefont {Wysmolek}}, \bibinfo {author}
  {\bibfnamefont {W.}~\bibnamefont {Strupinski}}, \bibinfo {author}
  {\bibfnamefont {C.}~\bibnamefont {Berger}}, \bibinfo {author} {\bibfnamefont
  {W.~A.}\ \bibnamefont {de~Heer}}, \bibinfo {author} {\bibfnamefont
  {G.}~\bibnamefont {Martinez}}, \ and\ \bibinfo {author} {\bibfnamefont
  {M.}~\bibnamefont {Potemski}},\ }\href@noop {} {\bibfield  {journal}
  {\bibinfo  {journal} {Phys. Rev. Lett.}\ }\textbf {\bibinfo {volume} {107}},\
  \bibinfo {pages} {216603} (\bibinfo {year} {2011})}\BibitemShut {NoStop}%
\bibitem [{\citenamefont {Pershoguba}\ and\ \citenamefont
  {Yakovenko}(2010)}]{PershogubaPRB10}%
  \BibitemOpen
  \bibfield  {author} {\bibinfo {author} {\bibfnamefont {S.~S.}\ \bibnamefont
  {Pershoguba}}\ and\ \bibinfo {author} {\bibfnamefont {V.~M.}\ \bibnamefont
  {Yakovenko}},\ }\href@noop {} {\bibfield  {journal} {\bibinfo  {journal}
  {Phys. Rev. B}\ }\textbf {\bibinfo {volume} {82}},\ \bibinfo {pages} {205408}
  (\bibinfo {year} {2010})}\BibitemShut {NoStop}%
\bibitem [{\citenamefont {Nakao}(1976)}]{NakaoJPSJ76}%
  \BibitemOpen
  \bibfield  {author} {\bibinfo {author} {\bibfnamefont {K.}~\bibnamefont
  {Nakao}},\ }\href@noop {} {\bibfield  {journal} {\bibinfo  {journal} {J.
  Phys. Soc. Jpn.}\ }\textbf {\bibinfo {volume} {40}},\ \bibinfo {pages} {761}
  (\bibinfo {year} {1976})}\BibitemShut {NoStop}%
\end{thebibliography}
%

\end{document}